\documentclass[referee]{aa} % for a referee version
\usepackage{graphicx}
\def\cl0048{$\mathrm{EIS\,0048}$}
\begin{document}
\titlerunning{Photometric Properties of Cluster Galaxies at $\mathrm{z\sim0.64}$}

\title{Photometric Properties of Galaxy Population in the Cluster
EIS\,0048-2942 at $\mathrm{z\sim0.64}$}

\subtitle{}

\author{F. La Barbera\inst{1} 
\and
P. Merluzzi \inst{2}
\and 
A. Iovino \inst{3}
\and
M. Massarotti \inst{2}
\and 
G. Busarello \inst{2}}

\offprints{F. La Barbera}

\institute{Universit\`a Federico II, Department of Physics, Napoli, Italy
email: labarber@na.astro.it
\and
I.N.A.F., Istituto Nazionale di Astrofisica 
Osservatorio Astronomico di Capodimonte, 
Via Moiariello 16, I-80131 Napoli, Italy\\
email: gianni@na.astro.it
\and
I.N.A.F., Istituto Nazionale di Astrofisica 
Osservatorio Astronomico di Brera, 
Via Brera 28, 20121 Milano, Italy}

\date{Received ; accepted }

\abstract{Deep photometric data in the V-, R-, I-, z- and K-bands for
the cluster of galaxies EIS\,0048-2942 are used to investigate the
properties of the galaxy populations at $\mathrm{z\sim0.64}$ in a
field of $\mathrm{2.5\times 2.5~Mpc^2}$. The sample of candidate
cluster members ($\mathrm{N = 171}$) is selected by the photometric
redshift technique and is complete up to $\mathrm{I=22.5}$. Galaxies
were classified as spheroids and disks according to the shape of the
light profile in the I-band, as parametrized by the Sersic index. In
both optical and NIR, spheroids define a sharp colour-magnitude
sequence, whose slope and zero points are consistent with a high
formation redshift ($\mathrm{z_f > 2}$).  The disk population occupies
a different region in the colour-magnitude diagram, having bluer
colours with respect to the red sequence. Interestingly, we find some
level of mixing between the properties of the two classes: some disks
lie on the colour-magnitude sequence or are redder, while some
spheroids turn out to be bluer. The spatial distribution of cluster
galaxies show a clumpy structure, with a main over-density of radius
$\mathrm{\sim0.5~Mpc}$, and at least two other clumps distant
$\mathrm{\sim1~Mpc}$ from the center. The various sub-structures are
mostly populated by the red galaxies, while the blue population has an
almost uniform distribution.  The fraction of blue galaxies in
EIS\,0048-2942 is $\mathrm{f_B=0.11 \pm 0.07}$. This is much lower
than what expected on the basis of the Butcher-Oemler effect at lower
redshifts.

   }

   \maketitle
%
%________________________________________________________________

\section{Introduction}

Since 1978 (Butcher \& Oemler~\cite{BO78a},~\cite{BO78b}), several
studies have been fruitfully undertaken in order to investigate the
galaxy evolution in clusters. 

Photometric and spectroscopic works found a systematic enhancement of
the fraction of blue galaxies in the cores of distant clusters up to
$\mathrm{z}\sim0.5$, known as Butcher -- Oemler (BO) effect
(e.g. Butcher \& Oemler~\cite{BO84}, Couch \& Sharples~\cite{CoS87},
Rakos \& Schombert~\cite{RaS95}, Margoniner et al.~\cite{MdC01},
Ellingson et al.~\cite{ELY01}), suggesting a strong increase in the
star formation of cluster galaxies with redshift (but see also
Koo et al.~\cite{KKN88}). It was also found that the blue galaxy
fraction increases with cluster-centric distance (e.g.~Kodama \&
Bower~\cite{KOB01}). Recent studies pointed out, however, that
selection criteria, as for instance the use of optically selected
samples, could affect significantly the determination of the BO effect
(e.g. Andreon \& Ettori~\cite{AnE99}, Fairley et al.~\cite{FJW02}).
For what concerns early-type galaxies, photometric studies at
different redshifts, showed that they appear to be a passively
evolving population, where the reddening of more massive galaxies with
magnitude (the colour-magnitude relation) is the result of a
mass-metallicity relation (e.g. Arago\'n-Salamanca et
al.~\cite{AEC93}; Kodama \& Arimoto 1997; Gladders et al. 1998;
Stanford et al.~\cite{SED98}; Kodama et al.~\cite{KAB98}, hereafter
KABA98).

Different studies have attempted to explain the observational
scenario.  Kodama \& Bower~(\cite{KOB01}) showed that the
colour-magnitude (hereafter CM) diagrams of cluster galaxies at
intermediate redshift can be explained by the continuous accretion of
field galaxies, whose star formation is inhibited by the hostile
cluster environment.  The origin of the blue galaxies has also been
investigated by using spectroscopic and morphological data. Barger et
al.~(\cite{BAR96}) and Poggianti \& Barbaro~(\cite{PoB96}) showed that
a significant fraction of galaxies in distant clusters could have
experienced a recent burst of star formation (but see also Abraham et
al. 1996).  Moreover, Couch et al.~(\cite{CBS98}) found that most of
the blue galaxies have the same properties of normal star forming
field spirals, while Dressler et al.~(\cite{DOC97}) showed that the
morphological content of distant clusters evolves significantly, the
fraction of S0s decreasing with redshift. Up to date, it is unknown
the origin of the mechanism that drives changes of star formation and
morphology when galaxies are accreted into the cluster, although it is
clear the effectiveness of detailed photometric and spectroscopic
studies in order to explain the scenario of galaxy evolution.

The cluster EIS\,0048-2942 (hereafter EIS\,0048) was detected by Olsen
et al.~(\cite{OSdC99a}) using the I-band galaxy catalogue of the ESO
Imaging Survey (EIS, Renzini \& da Costa 1997) and the cluster
redshift was estimated to be $\mathrm{z\sim0.6}$ by the matched filter
method (see Olsen et al.~\cite{OSdC99b}).  This candidate was also
identified in the EIS catalogue by Lobo et al.~(\cite{LIL00}) with an
estimated redshift $\mathrm{z\sim0.85}$ by a modified version of the
matched filter algorithm.  Later, a spectroscopic follow-up
(Serote-Roos et al.~\cite{SLI01}) confirmed the presence of a
structure at $\mathrm{z\sim0.64}$ with 15 concordant redshifts.

This work is the first of a series devoted to the detailed study of
this new galaxy cluster at $\mathrm{z\sim0.64}$, for which
multi-band photometry and spectroscopic data have been collected.

The paper consists {\rm of} two parts. The first part deals with data
reduction and sample selection, while in the second part we discuss
the photometric properties of the galaxy population. The reader
interested only in the second part may skip to Sect.~7.  In Sect.~2 we
present the photometric observations. Data reduction and photometric
calibration are described in Sects.~3 and~4 for VRIz and K-band,
respectively.  Aperture photometry is presented in Sect.~5, while
Sect.~6 deals with the photometric redshift estimates. In Sect.~7 we
introduce the galaxy shape classification.  The colour-magnitude
distributions are discussed in Sect.~8. The spatial distribution of the
cluster population and the BO effect are studied in Sects.~9 and~10,
respectively. A summary is given in Sect.~11.

In the following we assume $\Omega_m=0.3$, $\Omega_\Lambda=0.7$ and
$\mathrm{ H_0= 70~Km~s^{-1} Mpc^{-1} }$. With this cosmology the age
of the universe is $\mathrm{\sim13.5~Gyr}$, and the redshift
$\mathrm{z=0.64}$ corresponds to a look-back time of
$\mathrm{\sim6~Gyr}$.

\begin{figure*}
\begin{center}
\end{center}
\caption[]{ V-band FORS2 image of the cluster field. The rectangles
show the K-band ISAAC pointings, while the spatial scale is reported
in the lower--left corner.
\label{FIELDS}
}
\end{figure*}

\section{Observations}
New photometric observations of the cluster of galaxies EIS\,0048 were
carried out at the ESO Very Large Telescope (VLT) during two observing
runs in August 2001. All the nights were photometric with excellent
seeing conditions.

The data include VRIz and K-band imaging taken with FORS2 and ISAAC
instruments, respectively.  The VRIz images consist of a single
pointing of $6.8 \! \times \! 6.8~\mathrm{arcmin}^2$ for each band,
while for the K-band a mosaic of four pointings covers a total area of
$4.9 \! \times \! 4.9~\mathrm{arcmin}^2$.  Furthermore, we obtained a
mosaic of three FORS2 pointings in I-band by using the high resolution
(HR) observing mode. The reduction of the HR images will be presented
elsewhere.  In order to correct for the fringing patterns in the
I(SR)- and z-bands, we obtained five dithered exposures with a
dithering box of size $25''$.  For the K-band, we collected two sets
of 22 exposures for each pointing with $\mathrm{DIT=12~sec}$ and
$\mathrm{NDIT=6}$, by using a dithering box of size $15''$.  The
overlapping of the VRIz and K-band images is shown in
Fig.~\ref{FIELDS}, while the relevant information on the data are
summarized in Table~\ref{DATA}.
\begin{table}
\caption[]{Relevant information on the observations. Col. 1:
waveband. The symbols SR and HR denote the standard and high
resolution I-band images, respectively.  Col. 2: pixel scale. Col.
3: total exposure time for each pointing. Col. 4: full width half
maximum of the seeing disk. For the I(HR)- and K-bands, we
report the range for the various pointings.}
\label{DATA}
$$
%%\begin{array}{p{0.5\linewidth}l}
\begin{array}{cccc}
\hline
\noalign{\smallskip}
\mathrm{Band} & \mathrm{Scale} & \mathrm{T_{exp}} & \mathrm{Seeing} \\
&   ''/\mathrm{pxl} & \mathrm{ksec} & '' \\
\noalign{\smallskip}
\hline
\noalign{\smallskip}
{\rm V}     & 0.2  & 3.6 & 0.6 \\
{\rm R}     & 0.2  & 2.4 & 0.9 \\
{\rm I(SR)} & 0.2  & 1.2 & 0.8 \\
{\rm I(HR)} & 0.1  & 1.8 & 0.3 \! - \! 0.4 \\
{\rm z}     & 0.2  & 1.8 & 0.6 \\
{\rm K}     & 0.15 & 3.2 & 0.4 \! - \! 0.5 \\
\noalign{\smallskip}
\hline
\end{array}
$$
\end{table}

\section{VRIz data}

\begin{figure*}
\begin{center}
\end{center}
\caption[]{Correction of the fringing in the z-band image. Upper --
left and right panels show a portion of the z-band image before and
after the correction. The profiles of the average of three columns are
shown in the lower plots. Intensities are in units of the standard
deviation of the background, ${\rm \sigma_{BG}}$.
\label{ZFRING}
}
\end{figure*}

\subsection{Reduction}
The data reduction was performed by using IRAF~\footnote{IRAF is
distributed by the National Optical Astronomy Observatories, which are
operated by the Association of Universities for Research in Astronomy,
Inc., under cooperative agreement with the National Science
Foundation. } and Fortran routines developed by the authors.

The bias was corrected by using zero exposure images and the
prescan/overscan regions of the frames.  For each filter, the
scientific images were divided by a flat-field frame obtained by
combining twilight sky exposures. After this correction, however, low
frequency variations of $\sim5\%$ were still present across the
images. We obtained, therefore, a sky flat frame by median combining
the scientific exposures taken during the night.  In these frames,
residual objects were masked and a polynomial fit was applied.  The
division of the scientific frames by the fitting surfaces allowed to
reduce the low frequency variations across the chip to $\sim0.5\%$.

After this procedure, fringing patterns were noticeable in the R-, I-
and particularly in the z-band frames. For the R-band, we did not
apply any correction since the fringes, found only in the right side
of the chip, showed peak-to-peak variations in the final image of less
than half the r.m.s. of the background. For the I-band, fringing was
almost completely removed because of the dithering between the various
exposures. Since dithering did not remove the fringing pattern in the
z-band, we created a fringing model by filtering the relative sky flat
frame. The model was suitably scaled and then was subtracted from each
scientific image, as illustrated in Fig.~\ref{ZFRING}. This allowed to
reduce the peak to peak variations of the fringing pattern in the
final image from $\sim4$ times to $\sim0.5$ times the background
noise.

For each image, cosmic rays were detected by using the IRAF task
COSMICRAYS and a mask frame was produced including cosmic rays and
chip defects.  The images were corrected for the different airmass
with the relative extinction coefficients (see next section) and added
by using the corresponding masks with the IRAF task IMCOMBINE.

\begin{figure}
\includegraphics[angle=0,width=8cm,height=14.4cm]{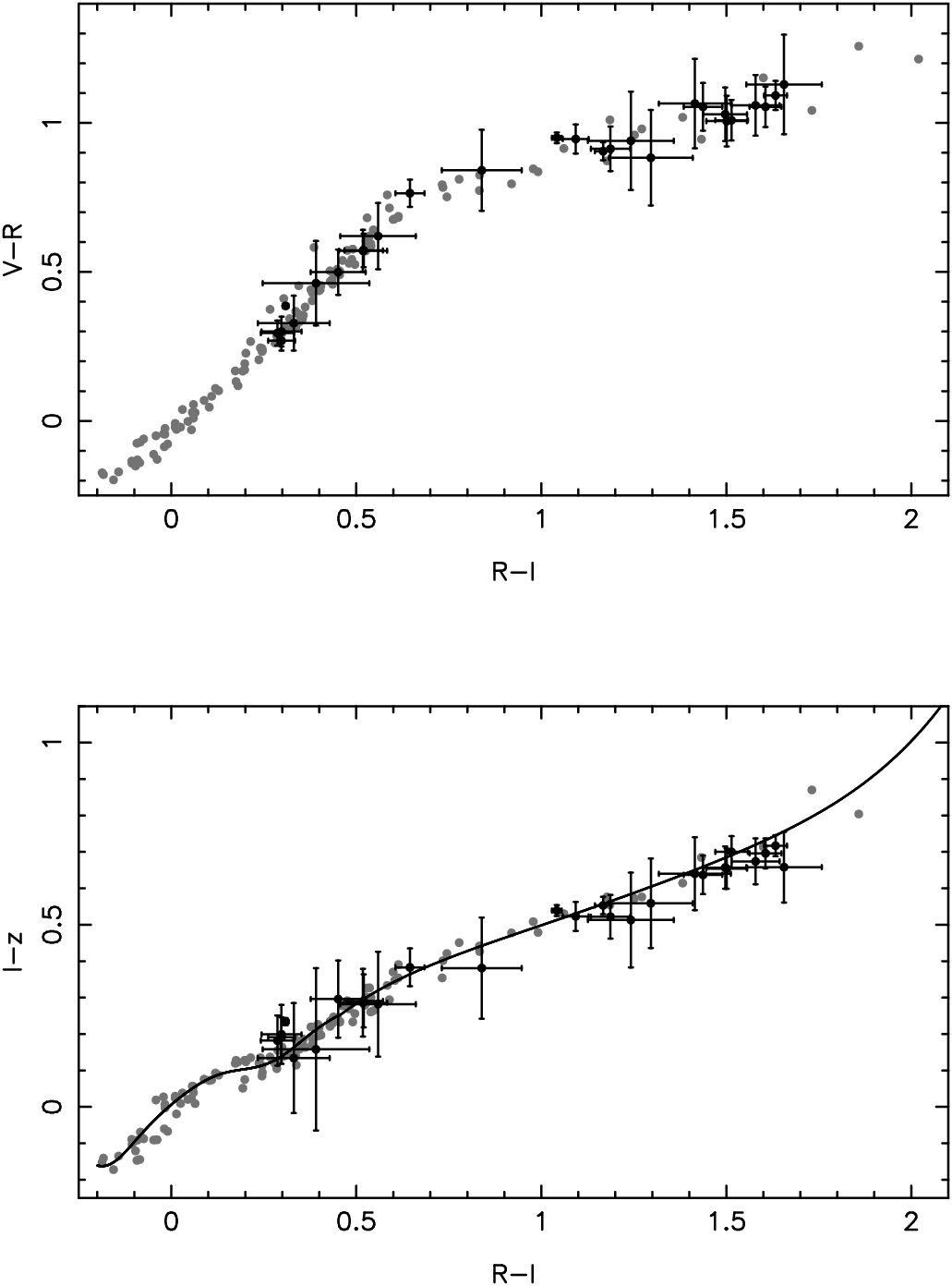}
\caption[]{Colour - colour diagrams of the stars in the cluster field
(black circles). Gray circles mark the synthetic colours for the
stellar spectra of Pickles~(\cite{PIC98}). The curve in the lower
panel is the polynomial fit to the locus of the Pickles stars,
that was used to match the observed and the synthetic distributions.
\label{ZCAL}
}
\end{figure}

\subsection{Photometric calibration}
The VRI images were calibrated into the Johnson-Kron-Cousins
photometric system by using comparison standard fields from
Landolt~(\cite{LAN92}) observed during each night. Instrumental
magnitudes of the standard stars were derived within an aperture of
$10 ''$ by running SExtractor (Bertin \& Arnouts~\cite{BeA96}). For
each waveband, we adopted the following calibration relation:
\begin{equation}
{\rm M = M' + \gamma \cdot C - A \cdot X + ZP }
\hspace{1cm} 
\label{calrel}
\end{equation} 
where M and C are magnitude and colour of the standard star, ${\rm M'}$
is the instrumental magnitude, $\gamma$ is the coefficient of the
colour term, ${\rm A}$ is the extinction coefficient, X is the airmass,
${\rm ZP}$ is the zero point.  For each waveband, the quantities ${\rm
\gamma}$, ${\rm A}$ and ${\rm ZP}$ were derived by a robust least
square fit to the data, except when standard stars with a sufficient
range of colours and/or airmasses were not available. In such case, we
adopted the value of $\gamma$ and/or ${\rm A}$ of the FORS2 standard
calibrations\footnote{\label{note_cal}
http://www.eso.org/observing/dfo/quality/FORS/qc/zero\-points/zeropoints.html.}
relative to the period of our observations. In Table~\ref{PHOTCAL}, we
report the fitting coefficients, the corresponding values of the FORS2
standard calibrations and the uncertainty on the photometric
calibration given by the r.m.s. of the residuals to the fit.  The
results of the fit are in very good agreement with the FORS2 standard
calibrations, with differences in the zero points smaller than ${\rm
0.03~mag}$.

To check the accuracy of the photometric calibration, we considered
the distribution in the (V-R,R-I) diagram of the stars present in the
cluster field. The observed colours were corrected for the
different seeing of the images as described in Sect.~5.  In
Fig.~\ref{ZCAL} (upper panel) we compare this distribution with that
obtained from the synthetic stellar spectra of the HILIB library
(Pickles~\cite{PIC98}). The synthetic colours were produced by using
the total efficiency curves of FORS2\footnote{ See the VLT exposure
time calculator at http://www.eso.org/observing/etc/.} and were
calibrated with respect to the Vega spectrum. The observed and the
synthetic distribution are in very good agreement, confirming the
accuracy of data reduction and calibration, and the photometric
quality of the observing nights.

Since no standard comparison fields were available for the z-band, we
adopted the following calibration method. Synthetic z magnitudes were
produced from the Pickles spectra and were calibrated by assuming
${\rm I-z=0}$ for the Vega spectrum. The zero point was obtained by
fitting the distribution in the (R-I,I-z) diagram of the stars in the
cluster field, corrected for seeing effects, to the locus of the
synthetic colours. The result of the fit is illustrated in
Fig.~\ref{ZCAL} (lower panel). The uncertainty on the zero point was
estimated with the bootstrap method by taking into account the
uncertainties on the colours of the stars. We obtained ${\rm ZP_z =
25.216 \pm 0.012}$. For the atmospheric extinction coefficient in the
z-band, we assumed ${\rm A_z = 0.08~mag/airmass}$, that is typical for
the Paranal site.
\begin{table*}
\caption[]{
\footnotesize
Results of the photometric calibration of the VRI data.
Col. 1: waveband. Col. 2: colour C used in Eq.~(\ref{calrel}).
Cols. 3, 4: colour term coefficients $\gamma$ and $\gamma'$. 
Cols. 5, 6: atmospheric extinction coefficients ${\rm A}$ and ${\rm A'}$.
Cols. 7, 8: zero points ZP and ${\rm ZP'}$ scaled to ${\rm 1 sec}$.
Col. 9: r.m.s. of the fitting residuals.
The quantities marked by a prime are the FORS2 standard calibrations
(see note~\ref{note_cal}).  The uncertainties refer to $1 \sigma$
standard intervals. Dots denote the quantities kept fixed in the
fits (see text).
}
\label{PHOTCAL}
$$
\begin{array}{ccccccccc}
\noalign{\smallskip}
\hline
   & {\rm C} & {\rm \gamma} & {\rm \gamma'} & {\rm A} &
{\rm A'} & {\rm ZP} & {\rm ZP'} & {\rm \sigma} \\ 
\noalign{\smallskip}
\hline 
\noalign{\smallskip} 
{\rm V} & {\rm V-I} & {\rm  .....}  & {\rm 0.010 \pm 0.005} &
{\rm 0.18 \pm 0.02} & {\rm 0.154 \pm 0.007} & {27.280 \pm 0.030} & {\rm 27.287
\pm 0.014} & 0.005 \\
{\rm R} & {\rm V-R} & {\rm .....}  & {\rm -0.018 \pm 0.011} & {\rm
0.10 \pm 0.01} & {\rm 0.114 \pm 0.007} & {27.420 \pm 0.013} & {\rm 27.450 \pm
0.015} & 0.013 \\
{\rm I} & {\rm V-I} & {\rm -0.058 \pm 0.004}  & {\rm -0.059 \pm 0.005} & {\rm
.....} & {\rm 0.053 \pm 0.007} & {26.474 \pm 0.003} & {\rm 26.484 \pm
0.037} & 0.018 \\
\noalign{\smallskip} \hline
\end{array}
$$
\end{table*}

\section{K-band data}

\subsection{Reduction}
The data reduction was performed by using ISAAC pipeline recipes
included in Eclipse, version 4.0, the InfraRed Data Reduction (IRDR)
software (see Sabbey et al.~\cite{SML01}) and Fortran routines
developed by the authors.

For both observing nights, we derived a dark frame by using the {\sl
dark} recipe of Eclipse. Since the dark variations between the two
nights were less than $1-2\%$, the dark frames were averaged and then
subtracted from all images.  As shown in Fig.~\ref{VPATT}, a double
ramp pattern along the column direction was noticeable in the
scientific images after dark subtraction. The effect was found to be
almost constant along the chip rows, varying from $\sim-5\%$ at the
rows where the detector starts to be read out (rows 1 and 513), to
$\sim3-5\%$ at the rows read last.  Since this pattern was not present
in the flat-field frame, it must be a residual additive signal
uncorrected by the dark subtraction.  As discussed in the ISAAC Data
Reduction Manual (version 1.5, Amico et al.~\cite{ACD02}), this effect
can be due to the dependece of the ISAAC infrared detector bias on
time and on illumination flux. To remove such an effect, we applied a
suitable correction immediately after dark subtraction. Each image was
collapsed by taking the median along the row direction, and a low-pass
filter was applied to the first 512 and to the last 512 samples of
this signal in order to remove high frequency variations due to the
noise and to the pixel to pixel response of the chip. The
one-dimensional frame obtained by this procedure (see
Fig.~\ref{VPATT}) was subtracted column by column from the original
image, removing the ramp pattern at better than $\sim1\%$.

\begin{figure}
\caption[]{Pattern remaining in the K-band images after dark
subtraction.  The gray line represents the signal after low-pass
filtering (see text).  
\label{VPATT}
}
\end{figure}

For the flat-field correction, we divided the scientific images by a
differential flat-field frame, obtained by two sets of twilight sky
exposures taken in the morning and in the evening,
respectively. Each twilight exposure was corrected for the odd-even
effect, particularly noticeable in the lower left quadrant (where it
amounted to $\sim2\%$ peak to peak variation). To this aim, we used
the {\sl oddeven} recipe of Eclipse.  For each set of twilight images,
a flat-field frame was obtained by subtracting high and low counts
exposures in order to correct for residual additive components not
removed by the dark subtraction.  The final flat-field was derived by
averaging the morning and evening twilight flats. This procedure
allows to minimize intensity variation within the array at
twilight. To test the accuracy of the flat-field correction, we
compared the magnitudes of each photometric standard star at the
different chip positions.  The differences were found to be
$\sim5-7\%$ ($1\sigma$ standard level).  To achieve higher accuracy,
we applied an illumination correction by retrieving from the ESO
archive the illumination frames closest in time to the date of our
observations. After illumination correction, the variation were found
to be $\sim1\%$, that we assumed as the final accuracy of the
flat-field.

The subsequent reduction included sky subtraction and image
combining, that were performed by a two-step procedure. First,
the images were processed by using the IRDR software. For each frame,
a sky image was obtained by a robust mean of the eight nearest
sequence exposures.  After sky subtraction, a mask image was created
for each frame by running SExtractor (Bertin \& Arnouts 1996) with the
checkimage OBJECTS option and was used to measure the dithering
offsets.  The first step coadded images were then obtained by
combining the exposures within each sequence.  At the second step, we
created an object mask for the coadded images obtained at the first
iteration by using SExtractor. These masks were expanded by a factor
of 1.5 and de-registered to the corresponding dither
exposures. Pixels with low and high counts in the flat-field were also
included into the masks in order to reject hot and cold
pixels. For each frame, the sky image was obtained by the average of
the six nearest exposures rejecting masked pixels. The images were
then coadded by using the IRAF task IMCOMBINE with a SIGCLIP algorithm
for cosmic ray rejection.

\subsection{Photometric calibration}
The photometric calibration of the ISAAC data was performed by using
standard stars from the list of Infrared NICMOS Standard Stars
(Persson et al.~\cite{PMK98}) and from the list of UKIRT standard
stars (see Hunt et al.~\cite{HMT98}).

The images were airmass-corrected by assuming an extinction
coefficient ${\rm A_K=0.08~mag/airmass}$, typical for the period of
our observations, and the total fluxes of the standard stars were
estimated within a circular aperture of diameter $8''$.  A small
systematic difference of $\sim0.05~{\rm mag}$ was found between the
zero points derived from the UKIRT and NICMOS standard stars. Such
result is not unexpected, since the UKIRT standards can be affected by
small systematic and/or random errors with respect to those of the
NICMOS list (see Sect. 5.2 of Persson et al.~\cite{PMK98}). The final
zero point was obtained, therefore, by considering only the NICMOS
standard stars, and was estimated to be $24.34\pm0.01~{\rm mag}$
(scaled to $1 {\rm sec}$ exposure time). This is in good agreement
with the value of the ISAAC pipeline (${\rm ZP\sim24.36}$), relative
to the period of our observations.

\section{Aperture photometry}
A catalogue of each image was produced by using the software
SExtractor (Bertin \& Arnouts~\cite{BeA96}). Since the images have
different seeing, particular care was taken in the choice of
deblending parameters in order that objects were deblended in the same
way for each band.  For each object, we obtained magnitudes within a
fixed aperture of diameter $2.3''$, corresponding to a physical size
of ${\rm \sim15~kpc}$ at ${\rm z=0.64}$, and magnitudes within an
aperture of diameter ${\rm \alpha \cdot r_K}$, where ${\rm r_K}$ is
the Kron radius (Kron~\cite{KRO80}). We chose $\alpha = 2.0$, for
which the Kron magnitude is expected to enclose $92 \%$ of the total
flux. The total magnitudes were computed by adding $\mathrm{0.08~mag}$
to the Kron magnitudes.

\begin{figure}
\includegraphics[angle=0,width=8.4cm,height=8.4cm]{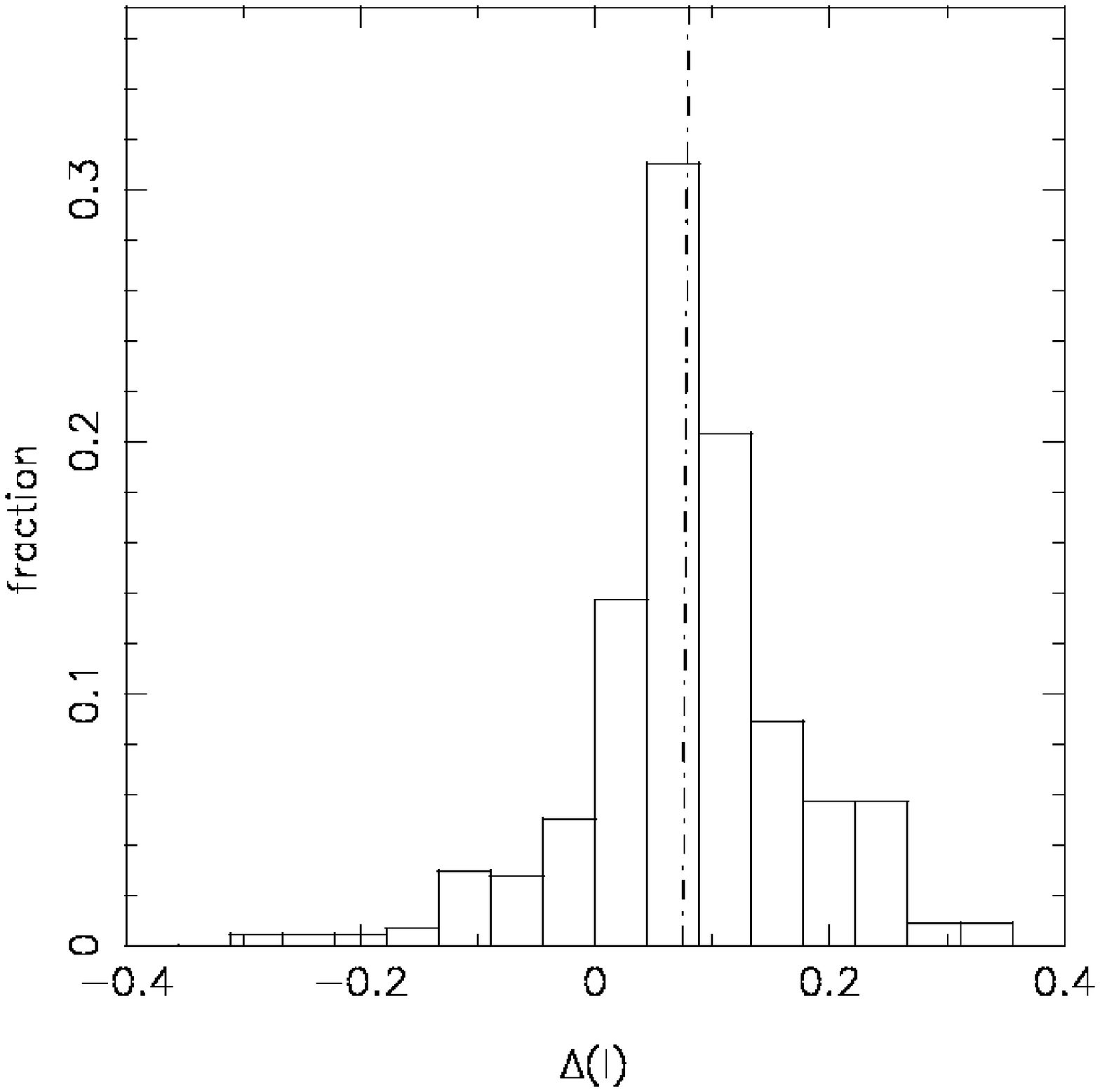}
\caption[]{Comparison of I-band high resolution and standard
resolution aperture magnitudes. Differences are in the sense SR - HR.
The dot -- dashed line marks the aperture correction estimated
by the method described in the text.
\label{APCOR}
}
\end{figure}

The catalogues were cross-correlated by deriving the coordinate
transformation between each image and the V-band image.  To this aim,
we proceeded as follows. A first list of matched objects was obtained
by using the IRAF task XYXYMATCH, with the triangle pattern
algorithm. This method has the advantage to be robust with respect to
geometric distortion effects in the images.  The final transformations
were obtained by fitting the coordinates of the objects in the
matching list with polynomials of order 5, which allowed to correct
for geometric distortion effects with an accuracy better than 0.5
pixel.  The derived coordinate transformations were used to match the
catalogues by using a nearest algorithm, producing a final list of
${\rm N=4868}$ sources detected in at least one band.  For the objects
with multiple K-band observations, the magnitudes turned out to be
fully consistent within the photometric errors, and the final
magnitudes were estimated by a weighted mean of the various
measurements.

Since the images have different seeing (cfr. Table~\ref{DATA}), the
derivation of the colour indices requires suitable aperture
corrections. To this aim, each image was convolved with a double
gaussian kernel, whose parameters were chosen interactively to match
the PSF of the image with the worst seeing (R band). Noise was added
in order to compensate for the smoothing introduced by the
convolution.  The aperture corrections were estimated by the robust
mean of the differences between the magnitudes of the objects in the
original and the convolved images, and amount to $\mathrm{0.05~mag}$
in V, $\mathrm{0.03~mag}$ in I, $\mathrm{0.06~mag}$ in z, and
$\mathrm{0.10~mag}$ in K.  The corresponding uncertainties were
estimated by the errors on the mean, and are $\mathrm{0.007~mag}$,
$\mathrm{0.013~mag}$, $\mathrm{0.02~mag}$, and $\mathrm{0.03~mag}$ in
V, I, z and K, respectively. These values were added in quadrature to
the photometric errors of the magnitudes. To test our procedure, we
compared the magnitudes of the sources in common between the standard
and high resolution I-band images, whose seeing values are
$0.3''$ and $0.8''$, respectively.  The magnitude differences are
shown in Fig.~\ref{APCOR} together with the aperture correction
derived as described above. The figure demonstrates the accuracy of
the adopted method.

The completeness in each band was estimated following the method of
Garilli et al.~(\cite{GAR99}), that consists in the determination
of the magnitude at which the objects start to be lost since they are
below the brightness threshold in the detection cell.  Details will
be given elsewhere (Massarotti et al. 2002, in preparation).  The
completeness magnitudes in V-, R-, I-, z- and K-band are 25.7, 25.0,
23.2, 22.5 and 21.2 respectively.
 
\section{Photometric redshifts}\label{zphot}
Photometric redshifts were estimated according to the Spectral Energy
Distribution (SED) fitting method (see Massarotti et al.~2001a,~b, and
references therein). In order to achieve a reasonable accuracy, we
considered galaxies with signal-to-noise ratio $\mathrm{S/N > 5}$ in
at least three bands. This sample contains ${\rm N = 633}$ galaxies,
of which ${\rm N=360}$ have K-band information, and is complete up to
the total magnitude ${\rm I_T=22.5}$.

\begin{figure}
\centering
\includegraphics[angle=0,width=8.5cm,height=8.5cm]{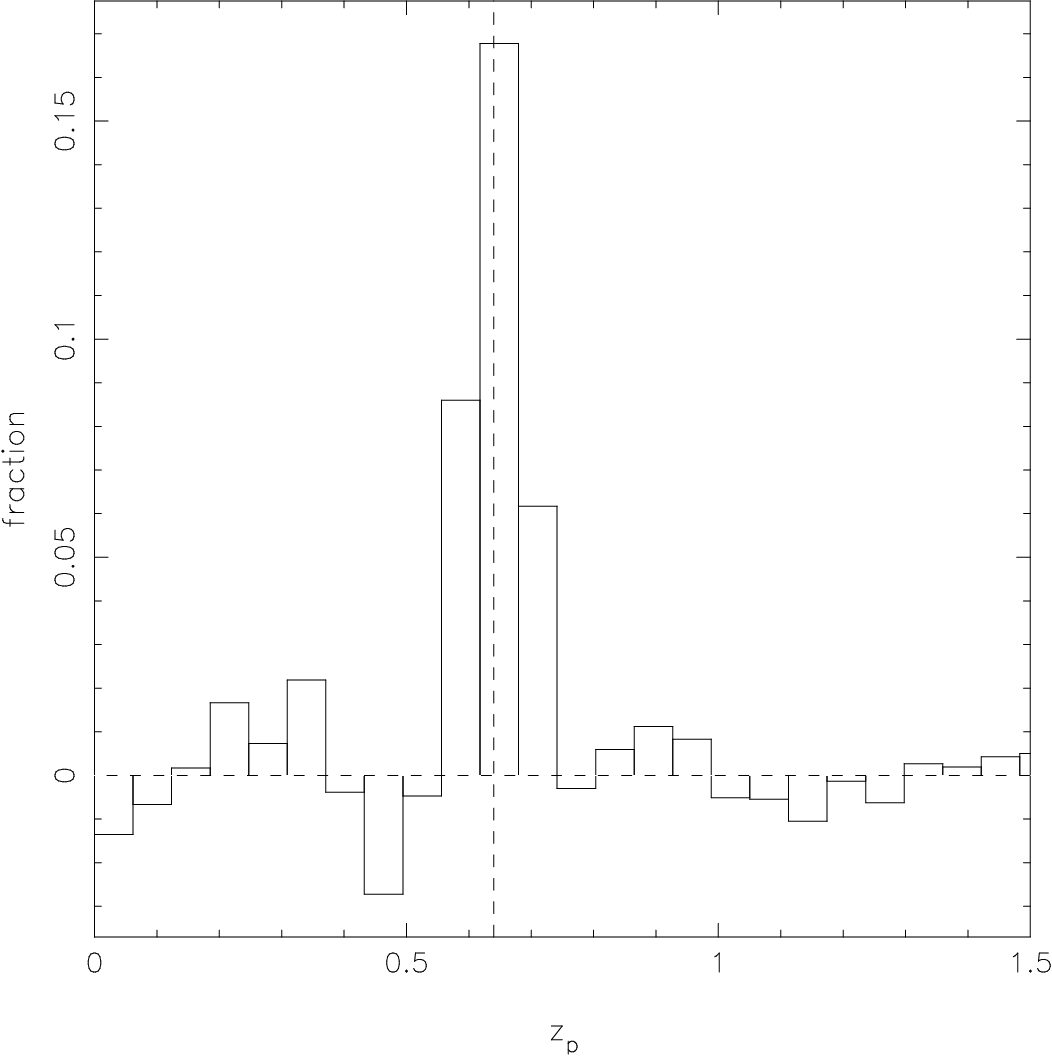}
\caption{Distribution of photometric redshifts $z_\mathrm{p}$ for
galaxies with $ 18.5 \le I \le 22.5$, after subtraction of the field
counts estimated by the VIRMOS survey data. The dashed line marks the
redshift 0.64. }
\label{HIST_ZF}
\end{figure}

The photometric redshifts were derived as described in Busarello et
al.~(\cite{BML02}), here we briefly outline the technique. Taking into
account the depth of our imaging, we looked for redshifts in the range
$z \in [0.0, 2.5]$ with a step of 0.01. Model galaxy spectra were
provided by the code of Bruzual \& Charlot~(\cite{BrC93}). The adopted
templates consist of models with a Scalo~(\cite{SCA86}) IMF and with
an exponential SFR $e^{-t/\tau}$.  We chose $\tau = 1, 4, 15$ Gyr, to
describe the colours of $E/S0$, $Sa/Sb$, and $Sc/Sd$ templates. To
allow for different metallicities of early-type galaxies, we
introduced $E/S0$ models with ${\rm Z/Z_\odot}=$0.2, 0.4, 1 and 2.5,
while template spectra evolution was followed in the time interval $t
\in [0.01,12.0]$ Gyr. To estimate the uncertainty ${\rm \Delta z_{p}}$
in the photometric redshift, we performed numerical simulations
by taking into account the uncertainties on magnitudes. Since the
K-band does not cover the whole cluster field, we investigated the
possible bias on $\mathrm{z_{p}}$ for the galaxies without K-band
data. To this aim, we compared the photometric redshifts obtained for
the sources with $\mathrm{I_T < 22.5}$ (a) by taking into account and
(b) by ignoring the K-band.  
{\bf The number of galaxies which were found to be cluster members 
in case (a) and no cluster members in case (b), and viceversa, turned
out to be negligible ($ < 5 \%$).}

We defined a galaxy as a cluster member when the photometric redshift
is in the range ${\rm z_{p} \in [0.54,0.74]}$.  However, to obtain a
reliable sample of cluster members via photometric redshifts it is
crucial to take into account possible degeneracies of template colours
at different redshifts. By inspecting colours of GISSEL98 templates,
we found that some degeneracy exists between colours of {\it blue}
$Sc/Sd$ galaxies at the cluster redshift ($\mathrm{V-R=0.94}$,
$\mathrm{R-I=0.94}$, $\mathrm{I-K=2.62}$ with $\mathrm{t_{form}= 12~
Gyr}$, $Z = Z_\odot$) and those of {\it red} $E/S0$ galaxies at
$z\sim0.3$ ($\mathrm{V-R=1.04}$, $\mathrm{R-I=0.82}$,
$\mathrm{I-K=2.80}$ with $\mathrm{t_{form}= 12~Gyr}$, $Z = Z_\odot$).
To correct for this effect, we also considered as cluster members the
objects having $z \in [0.3,0.4]$ and ${\rm \Delta z_{p}}$ consistent
with the cluster redshift. We found that ${\rm N=19}$ galaxies satisfy
this criterion. Most of these galaxies ($\sim75\%$) turn out to
be disk dominated objects and therefore their late spectral type is
also supported by the shape of the light profile (Sect.~7). The final
list of cluster members brighter than ${\rm I=22.5}$ consists in ${\rm
N = 171}$ galaxies.

To gain insight into the contamination by field galaxies in our
sample, we used the VIRMOS preparatory photometric survey as the
control sample. These data, kindly provided by the VIRMOS Consortium
(Le F\`evre et al. 2002, in preparation), are actually the proper
dataset for estimating the field contamination in our sample because
of their large field, similar filters and high depth).  In particular,
to estimate the redshift distribution for field galaxies we used the
VIRMOS catalogues (VRIK, McCracken et al. 2002, in preparation)
relative to a 150~arcmin$^2$ field with available K-band photometry.
It has to be noticed that the photometric redshifts were obtained for
the field and the cluster by using a similar photometric baseline and
the same procedure.  The distribution of photometric redshifts, after
subtraction of the field counts, is shown in Fig.~\ref{HIST_ZF} for
the objects with $ 18.5 \le$ I $\le 22.5$, corresponding to the
magnitude range of cluster galaxies up to the completeness limit of
the sample with photometric redshifts. It is dominated by the peak
around the cluster redshift.  The FWHM of the peak is $\sim0.1$ and
is due to the intrinsic redshift distribution of the cluster members
and to the errors on photometric redshifts. The redshift distribution
of VIRMOS galaxies is in full agreement with that relative to the
cluster field outside of the peak at $\mathrm{z}\sim0.64$, thus
confirming the effectiveness of our selection procedure and the
reliability of the VIRMOS control field.  In the field of EIS\,0048 we
found $\mathrm{N}=140\pm12$ galaxies with $\mathrm{z}<0.5$ and
$\mathrm{N}=46\pm7$ galaxies with $0.8<\mathrm{z}<1.8$, while in the
same redshift ranges the number of galaxies predicted on the basis of
the control sample are $\mathrm{N}=141.5\pm6.6$ and
$\mathrm{N}=43\pm4$ respectively. According to the VIRMOS data,
$\mathrm{N}=48\pm4$ ($25\%$) galaxies of the cluster sample are
expected to be field galaxies.

\begin{figure}
\centering
\includegraphics[angle=0,width=8.5cm,height=8.5cm]{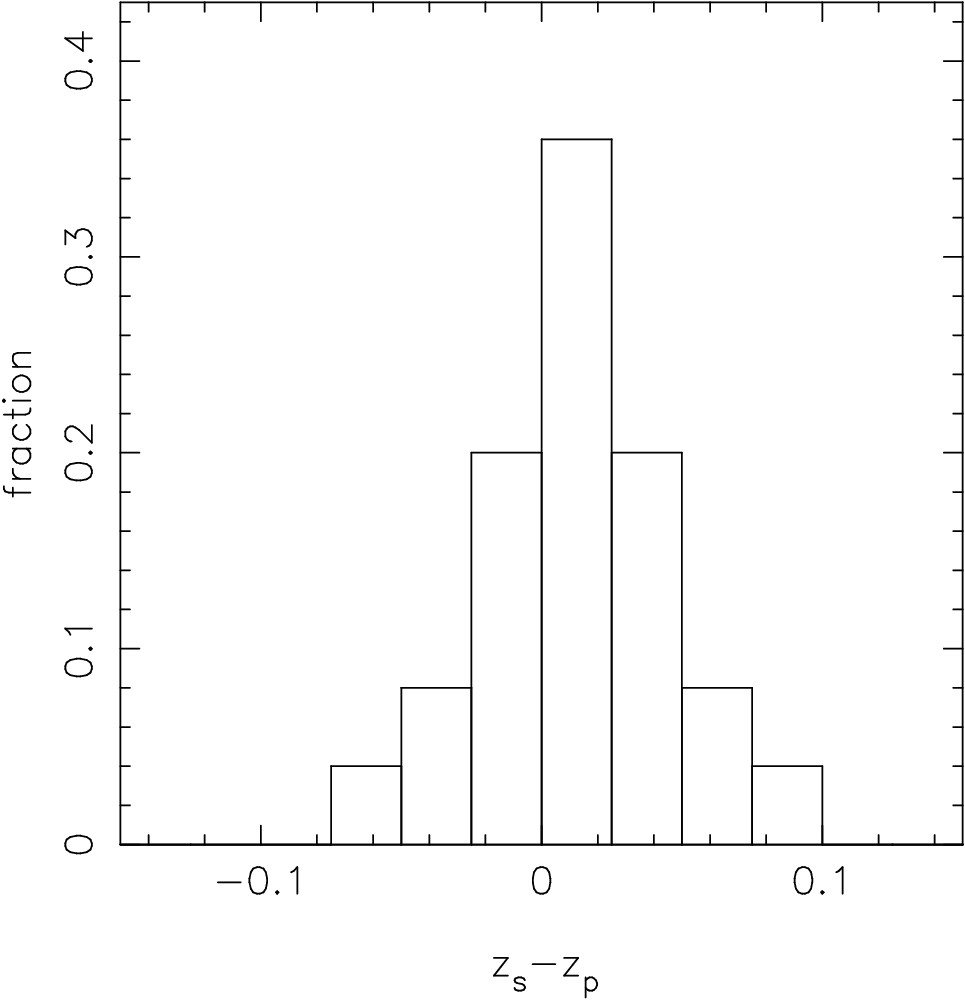}
\caption{Comparison of spectroscopic $z_\mathrm{s}$ and photometric
redshifts $\mathrm{z_{p}}$ for 25 spectroscopically confirmed
cluster members.  }
\label{photspec}
\end{figure}

In Fig.~\ref{photspec} we compare photometric ($z_\mathrm{p}$) and
spectroscopic ($z_\mathrm{s}$) redshifts for a sub--sample of 25
cluster galaxies (Busarello et al. 2002, in preparation, and C. Lobo,
private communication). 
We notice that all the spectroscopically confirmed cluster members
satisfy the selection criterion previously discussed.

\section{Galaxy shape classification}
Although a proper morphological classification requires the high
quality of the HST images, relevant information on the properties of
the light distribution of distant galaxies can be obtained by using
ground based data (see La Barbera et al.~\cite{LBM02}, hereafter
LBM02, and references therein). We used the procedure described in
LBM02 to model the surface brightness distribution of galaxies in the
cluster field, by fitting Sersic models convolved with the PSF of our
images. This analysis will be presented in detail in a forthcoming
paper.

For the present work, we will use the information on the shape of the
galaxy light profiles as parametrized by the Sersic index n. We fitted
Sersic models both to the standard resolution (SR) and to the high
resolution (HR) I-band images.  Accurate structural parameters for the
galaxies at the redshift of EIS\,0048 can be obtained from the very
good seeing of the HR images (La Barbera et al. 2002, in preparation).
Nevertheless, we found that a reliable separation between galaxies
with low n value (disks) and galaxies with a higher value of the
shape parameter (spheroids) can be also done with our SR images,
allowing us to perform the galaxy shape classification in the whole
observed field.  In fact, only $\sim8\%$ of the galaxies brighter than
${\rm I_T = 22}$ have a different classification in the HR and the SR
images, while the number of mismatches increases to $\sim17\%$ at
${\rm I_T\sim22.5}$. Moreover, for the galaxies with discordant
classification we did not find any significant systematic effect:
$\sim10\%$ of the galaxies defined as disks on the basis of the HR
data are classified as spheroids in SR, and vice-versa for the
remaining $\sim7\%$.

We chose ${\rm n=2}$ to discriminate between the two classes of
galaxies. As shown by numerical simulations (see van Dokkum et
al.~\cite{vDF98}), this criterion corresponds to separate objects with
a low bulge fraction ($<20 \%$) from galaxies with a larger bulge
component. Our final sample of galaxies with shape information
consists of all the ${\rm N=171}$ cluster members brighter than ${\rm
I_T=22.5}$.
 
\section{Colour-magnitude relations}
The optical (V-I) and NIR (V-K) colour-magnitude diagrams for the
objects in the cluster field are shown in Fig.~\ref{CM}. The black
dots denote objects brighter than $\rm{I_T}=22.5$ defined
as cluster members according to their photometric redshift (see
Sect.~\ref{zphot}), while different symbols are used to mark disks and
spheroids. At this magnitude limit, we have shape classification for
all the cluster members.
% The cluster sample is complete up to $\rm{I_T} \sim 22.5$
% (${\rm N = 171}$) and $\rm{K_T} \sim 19$ (${\rm N = 79}$) respectively
% in the V-I and V-K diagrams.  At the same limits, we have shape
% classification for all the cluster members.

\begin{figure}
\caption[]{ Colour-magnitude diagrams for the galaxies in the cluster
field (gray circles). Black symbols denote objects with $\rm{I_T}
< 22.5$ defined as cluster members on the basis of their photometric
redshift. Circles denote the spheroids, while disks are marked by
crosses. The dashed lines are the best fits to the CM while the dotted
lines limit the $\pm \sigma$ intervals.  The error bars in the left
corners of the plots indicate the mean uncertainties on colours and
magnitudes at $\mathrm{I_T\sim19.7}$ and $\mathrm{I_T\sim22.2}$ for
the upper plot and at $\mathrm{K_T\sim17}$ and $\mathrm{K_T\sim19}$
for the lower plot.
\label{CM}
}
\end{figure}

\subsection{Distribution in the CM planes}
\label{DCM}
Spheroids define sharp CM relations both in the (V-K,K) and in the
(V-I,I) planes. The sequences extend for at least 3 magnitudes, up to
the completeness limits of our samples, and are defined in the plots
by the loci enclosed within the dotted lines (see next section for
details). It is interesting to notice that 43 spheroids are
significantly bluer with respect to the $V-I$ red sequence, among them
24 are bluer in the range $[-0.4,-0.2]~\mathrm{mag}$. In particular,
we find six blue spheroids at very bright magnitudes ($\mathrm{I_T} <
20.2$ and ${\rm 1.9<V-I<2.2}$). Two of these have also K-band
photometry and lie on the red locus in the (V-K,K) plane.  We verified
that the classification of these blue spheroids is the same in all our
images, and therefore it is not affected by a possible uncertainty on
the Sersic index. Most of the disks (43) are bluer with respect to the
optical red locus, although $10$ galaxies do not show significant
deviations and other two have redder colours.  Four of these 12
galaxies have K-band photometry and are located above the V-K CM
relation.

Since our sample is selected by the photometric redshift technique,
some considerations are needed about the contamination by field
objects in our diagrams. On the VIRMOS data, in the (V-I,I) plane we
expect $39.5\pm 3.5$ blue galaxies, $8 \pm 1.6$ red sequence objects
and $0.6 \pm 0.4$ redder galaxies.  These numbers indicate that the
presence of blue spheroids and red disks is not due to the field
contamination but that at least some of these objects are real cluster
members.  Moreover, on the basis of the comparison field sample, the
cluster membership of the bright blue spheroids is highly
significant. In fact, in the range of magnitude and colours of these
galaxies, we expect that only $0.5 \pm 0.5$ objects could be field
contaminants.

\subsection{Slope and zeropoint of the CM relations}
\label{DCM2}
We study the CM relation of EIS\,0048 by comparing our results with
the models of KABA98.  We performed a linear regression of the V-K
and V-I colour-magnitude sequences by considering the following
relations:
\begin{eqnarray}
{\rm V-I \, } & = & {\rm  A_{VI} + B_{VI} \cdot I}\\ 
{\rm V-K } & = & { \rm A_{VK} + B_{VK} \cdot K}.
\label{CMREL}
\end{eqnarray}
The fits were done by minimizing the bi-weight scatter of the
residuals (see Beers, Flynn \& Gebhardt~\cite{BFG90}), and the
algorithm was modified in order to weight each point with the
corresponding uncertainty on the photometric redshift estimate. Only
objects within the completeness limit of our photometry were
considered. This procedure allows to minimize the effect of outliers
and of objects with poor photometric information and/or cluster
membership. We also verified that the fitting results do not change by
excluding disks from our sample. The uncertainties on the fitting
parameters were obtained by numerical simulations of the distributions
in the CM planes. For the optical relation, we obtained ${\rm A_{VI}=
3.95 \pm 0.30~mag}$ and ${\rm B_{VI}= -0.073 \pm 0.015 }$, with a
r.m.s.  of the residuals that amounts to ${\rm \sigma_{VI} = 0.10 \pm
0.03~mag}$.  For the NIR, due to the smaller sample size, the relative
uncertainty on the CM slope is large: ${ \rm B_{VK}=-0.15 \pm
0.08}$. Therefore, we chose to repeat the fit by using a fixed value
of the slope ${\rm B_{VK}=-0.11}$, that is typical for the spheroids
in clusters at the same redshift (see Fig.~4 in KABA98).  This gives
${\rm A_{VK}=7.75 \pm 0.05~mag}$, with r.m.s. of the residuals ${\rm
\sigma_{VK} = 0.14 \pm 0.06~mag}$. The fits to the CM relation are
shown in Fig.~\ref{CM} along with the relative $3~\sigma$ intervals.
We denote an object as a blue galaxy if it lies more than
$3~\sigma$ below the (V-I,I) CM fit, as a red galaxy otherwise.

\begin{figure*}
\begin{center}
\end{center}
\caption[]{ I-band image of the cluster field. The map of the counts
density is represented by the gray contours. The grey intensity is
proportional to the density value. Triangles and circles mark disks
and spheroids respectively.
\label{DMAP}
}
\end{figure*}

By considering the previous results, we see that the dispersions
around the optical and NIR CM relations can be fully explained by the
typical uncertainty on galaxy colours (${\rm \sim0.1~mag }$,
cfr. Fig.~\ref{CM}), and that, therefore, the intrinsic scatter of the
relations is very small. Unfortunately, due to the field contamination
and to the uncertainty on morphological selection, the quoted errors
on the dispersions are too large to obtain an accurate quantitative
estimate of the intrinsic scatter. The fitting coefficients of the
Eqs.~(\ref{CMREL}) can be compared with the predictions of the KABA98
models relative to the ${\rm V_{555}-I_{814}}$ vs. ${\rm I^T_{814}}$
and the ${\rm V-K}$ vs ${\rm K^T}$ CM relations considered by the
authors (see Figs.~4,~5 in KABA98), where ${\rm V_{555}}$ and ${\rm
I_{814}}$ are the magnitudes of the HST filters.  By comparing the
value of ${\rm B_{VI}}$ with the values shown in Fig.~4 (right middle
panel) in KABA98, we see that our estimate is fully consistent with the
prediction of a pure metallicity sequence with an old epoch (${\rm
z_f=4.5}$) of galaxy formation. Pure metallicity models, with a more
recent formation epoch, ${\rm z_f=1.7}$ and ${\rm z_f=1.2}$, are
discarded at $1.5 \sigma$ and $2.0 \sigma$ confidence level
respectively.  The zero points of the CM relations can be compared
with the values shown in Fig.~5 in KABA98. To this aim, we referred the
${\rm A_{VI}}$ and ${\rm A_{VK}}$ coefficients to the magnitudes
$\mathrm{K_T}=-25.5$ and $\mathrm{I_T}=-22$,
by using the same cosmology adopted by
KABA98 (${\rm H_0 = 50~Km~s^{-1} Mpc^{-1}}$, $\Omega_m = 0.3$,
$\Omega_{\Lambda}=0.7$) and by estimating ${\rm K}$ corrections from
Poggianti~(\cite{Pog97}). For the NIR, we obtain a corrected value of
the zero point ${\rm A^c_{VK}=5.58 \pm 0.05~mag}$, that is consistent
with the models of KABA98 having ${\rm z_f \ge 1.7}$, while younger
models are not consistent with our data. For the optical zero point,
we obtain ${\rm A^c_{VI} = 2.33 \pm 0.03~mag}$ that is about 0.2~mag
below the value relative to the model with ${\rm z_f\sim4.5}$. We
notice, however, that this difference can be affected by the
uncertainty on the zero point of the HST photometry (see e.g. Holtzman
et al.~\cite{HBC95}), making this disagreement not significant. In
fact, the comparison of the ${\rm R-K}$ CM with KABA98 gives the same
results obtained for the ${\rm V-K}$ relation.  Moreover, we notice
that the ${\rm V-I}$ colour predicted by the GISSEL98 synthesis code
(see Bruzual \& Charlot~\cite{BrC93}) for an E template, with solar
metallicity and with ${\rm z_f} = 4.5$, is ${\rm 2.34~mag}$, in
full agreement with the value of ${\rm A^c_{VI} }$.

\section{Spatial distribution of cluster populations}
To further investigate the properties of the blue and red galaxy
populations, we analyzed their spatial distribution in the cluster
field. In Fig.~\ref{DMAP}, we show a map of the number density of the
objects defined as cluster members according to their photometric
redshifts. Cluster galaxies show an evident clumpy distribution: the
peak of the density is associated to a main structure having projected
radius $\sim 1.0'-1.5'$ ($\mathrm{\sim0.5~Mpc}$ at ${\rm z=0.64}$),
while at least two secondary clumps of galaxies are found at a
distance of $\sim3.0'$ ($\mathrm{ \sim1.2~Mpc}$) from the main
peak. Although we cannot quantify the number of spheroid and disks
that are expected to be field objects, it is quite remarkable that the
various structures are populated, mostly, by galaxies with a more
concentrated shape, while disks are found preferably in the low
density regions.

\begin{figure}
%\vspace{2cm}
%\hspace{4cm}
%\includegraphics[angle=0,width=8.5cm,height=18.9cm]{H4104F10.ps}
\caption[]{ Map of the counts density for the red and blue cluster
populations (upper and lower panels). The grey scale is the same for
both panels.
\label{DMAPS}
}
\end{figure}

By using the separation between red and blue galaxies discussed
in Sect.~\ref{DCM2}, we constructed the density maps relative to the
two classes of objects (see Fig.~\ref{DMAPS}).  The clumps of galaxies
defining the cluster structure are evident in the plot relative to the
red population, for which we expect a very low field contamination
($\sim7\%$). With the exception of a possible clump near the border of
the field, the density map of the blue population does not show any
particular structure.  This is partly due to the field contamination,
but it could also indicate that cluster blue galaxies do not share the
same spatial distribution of the red ones. In fact, if we construct
the density map of the blue galaxies, by weighing each object with the
relative luminosity, we find that the clumps present in the lower
panel of Fig.~\ref{DMAPS} outside the region corresponding to the
cluster central structure, become the dominant features in the
plot. We point out, however, that a spectroscopic survey in the field
is needed in order to definitely exclude the contamination effect.

\section{Butcher-Oemler effect}
In order to derive the fraction of blue galaxies ${\rm f_B}$ in the
cluster EIS\,0048, we followed the procedure of Butcher \&
Oemler~(\cite{BO78a},~\cite{BO78b}). We computed the fraction of
galaxies that are ${\rm 0.2~mag}$ bluer than the red sequence in the
rest-frame (B-V,V) diagram within a magnitude limit ${M_V = -20}$, and
that are enclosed within ${R_{30} }$, the radius containing $30\%$ of
the total galaxy population.

\begin{figure}
\caption[]{ Density profile of the galaxies defined as cluster
members. Error bars denote Poissonian uncertainties. Density have been
arbitrarily scaled in order to have a density of $\mathrm{1~gal/arcmin^2}$
for the field (dashed horizontal line). The curves show the models
fitted to the density profile (see upper-right).
\label{DPROF}
}
\end{figure}

In Fig.~\ref{DPROF}, we show the density profile of the cluster
members derived by considering circular coronae centered on the peak
of the density map (see Fig.~\ref{DMAP}) with the suitable correction
for the rectangular geometry of our field. The density counts relative
to the cluster core decrease gradually up to a distance of ${\rm
\sim2~arcmin}$, while the substructures of Fig.~\ref{DMAP} are shown
by the increase of the profile between ${\rm R=2'}$ and ${\rm
R=4'}$. At large distances, the density is very close to the value
expected in the field, showing that the estimate of field
contamination is reliable.  To derive the value of ${\rm R_{30}}$, we
need to predict the fraction of cluster galaxies ${\rm f_O}$ expected
out of the observing field. To this aim, we used an extrapolation of
the density profile, by fitting the density counts of the cluster core
(${\rm R<2' }$) by different models. We considered the models adopted
by Adami et al.~(\cite{AMK98}) to analyze a large sample of nearby
clusters: the generalized Hubble and King profiles, the generalized
NFW (see Navarro, Frenk \& White~\cite{NFW95},~\cite{NFW96}) and de
Vaucouleurs models. The fitted profiles were integrated from ${\rm
R=4'}$ up to an Abell radius (${\rm 3 h_{50}^{-1} Mpc}$), giving ${\rm
f_O = 20 \pm 5 \%}$. This value was used to calculate ${\rm R_{30}}$
and the concentration parameter ${\rm C= \log (R_{60}/R_{20})}$, where
${\rm R_p}$ is the radius that encloses ${\rm p\%}$ of the total
number of cluster galaxies. The same procedure was repeated by
extrapolating the density profile out of ${\rm R}=2'$, without
considering the substructures present between ${\rm R}=2'$ and ${\rm
R}=4'$. In Table~\ref{CPAR} we report the values of ${\rm C}$ and ${\rm
R_{30}}$ estimated for case 1 (R$<2'$) and for case 2 t(R$<4'$), and
the corresponding uncertainties, obtained by taking into account the
Poissonian error on our number counts and the uncertainty on the
density of field counts.  As it was expected, the value of ${\rm
R_{30}}$ is smaller for case 1, and ranges between ${\rm
R_{30}}\sim1'$ and ${\rm R_{30}}\sim1.5'$, while the value of ${\rm
C}$ varies from $\sim0.43$, that is typical of a not particurarly
compact cluster, to $\sim0.5$, that is closer to that of a more rich,
concentrated structure (see Butcher \& Oemler~\cite{BO78b}). To check
our estimate of ${\rm R_{30}}$ and ${\rm C}$, we repeated the
calculation by considering only the red cluster members. We obtained
${\rm R_{30}} = 1.2 \pm 0.1$ and ${\rm C} = 0.53\pm0.02$ for case 1,
while ${\rm R_{30}} = 1.5 \pm 0.19$ and ${\rm C} = 0.44\pm0.013$ for
case 2.  These values are fully consistent with those reported in
Table~\ref{CPAR}.

\begin{table}
\caption[]{Standard parameters for the computation of the BO effect.
See text for details.
\footnotesize
\label{CPAR}
}
$$
\begin{array}{lcccc}
\noalign{\smallskip}
\hline
 & {\rm C} & {\rm \Delta C} & {\rm R_{30}} & {\rm \Delta R_{30}} \\
 & &  & (') & (') \\
\hline
 {\rm 1 (R<2')} & 0.50 & 0.02 & 1.0 & 0.07\\
 {\rm 2 (R<4')}  & 0.43 & 0.02 & 1.5 & 0.17\\
\hline
\end{array}
$$
\end{table}

\begin{figure}
\includegraphics[angle=0,width=8.5cm,height=8.5cm]{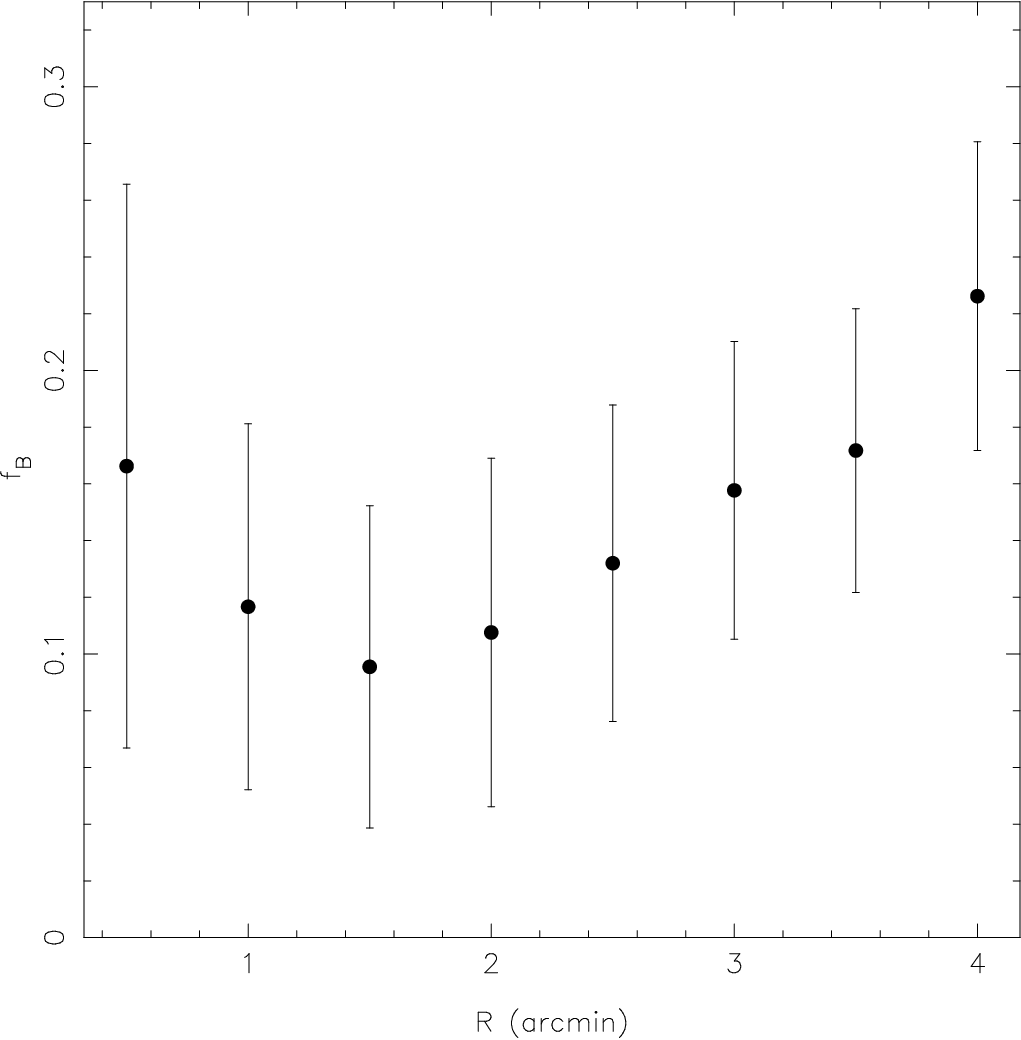}
\caption[]{ Fraction of blue galaxies within R, where R is the
cluster-centric distance.
\label{BOEFF}
}
\end{figure}

The blue fraction ${\rm f_B}$ was computed by using the ${\rm V-I}$
colour index, that is very close to ${\rm B-V}$ rest-frame at the
cluster redshift. In order to convert ${\rm V-I}$ colour into ${\rm
B-V}$ rest-frame, we used the synthesis code of Bruzual \& Charlot
(1993) to construct galaxy templates with different age, metallicity,
star formation rate, and dust content, and we computed for each of
them ${\rm V-I}$ and ${\rm B_r-V_r}$ colours, where $\mathrm{B_r}$ and
$\mathrm{V_r}$ are magnitudes in the redshifted B and V filters. The
relation between the two colours was found to be described very well
by a polynomial of degree four, with a very small r.m.s. of $0.025
{\rm~mag}$.  The same procedure was adopted to convert the ${\rm I}$
magnitudes into rest-frame V-band magnitudes.  The value of ${\rm
f_B}$ was computed by subtracting the fraction of blue galaxies
obtained for the cluster sample to that expected for the field.  The
calculation was performed by considering only galaxies brighter than
${\rm M_V}=-20$, that corresponds to ${\rm I}\sim22.5$ at ${\rm
z=0.64}$. We notice that this corresponds to the completeness limit of
our cluster sample. The uncertainty on ${\rm f_B}$ was estimated by
taking into account Poissonian and photometric errors of both cluster
and field samples.  In Fig.~\ref{BOEFF}, we show the trend of ${\rm
f_B}$ versus the cluster-centric distance.  At distances between ${\rm
R=1'}$ and ${\rm R=1.5'}$, that enclose our estimate of ${\rm R_{30}}$
(cfr. Table~\ref{CPAR}), we obtain ${\rm f_B=0.11 \pm 0.07}$. We notice
that this result is very robust with respect to our computation
of ${\rm R_{30}}$: ${\rm f_B}$ is, in fact, practically constant
between ${\rm R}=1'$ and ${\rm R=}2'$.

The blue galaxy fraction of the cluster EIS\,0048 can be compared with
the value predicted by the standard BO effect at ${\rm z=0.5}$, ${\rm
f_B=0.25}$, and the values obtained by recent works, who have
attempted to compute the BO effect at ${\rm z>0.5}$. In particular,
the BO effect has been investigated at ${\rm z\sim0.8}$ by van Dokkum
et al.~(\cite{vDF00}), who found ${\rm f_B=0.22 \pm 0.05}$, and by
Rakos \& Schombert~(\cite{RaS95}), who estimated a very high blue
galaxy fraction, ${\rm f_B\sim0.8}$. In agreement with van Dokkum et
al.~(\cite{vDF00}), our results do not indicate a high increase in the
BO effect at ${\rm z > 0.5}$. The cluster EIS\,0048 at ${\rm z=0.64}$
shows a quite low blue galaxy fraction, about one half of the standard
BO effect at ${\rm z=0.5}$. By looking at Fig.~\ref{BOEFF}, it is also
interesting to notice that the value of ${\rm f_B}$ seems to increase
at distances greater than ${\rm R=2'}$ up to $\sim23\%$ at ${\rm
R=4'}$. This would be consistent with what found by previous studies
at lower redshift (e.g. Kodama \& Bower~\cite{KOB01}, Fairley et
al.~\cite{FJW02}), and can be explained as a consequence of accretion
of field galaxies into the cluster structure.

\section{Summary.}
We have studied the properties of the galaxy populations in the
cluster \cl0048 \, at $z=0.64$, by using a large photometric baseline
including V-, R- I-, z- and K-band data. Cluster members have been
selected by using the photometric redshift technique, producing a
final list of $\mathrm{N=171}$ galaxies complete up to
$\mathrm{I=22.5}$. To estimate the field contamination, we used as
control sample the VIRMOS preparatory photometric survey, for which
photometric redshifts were estimated by using the same procedure
adopted for the cluster sample. Out of the $\mathrm{N=171}$ candidate
members, N=48 objects are expected to be field contaminants.  Cluster
galaxies have been classified into disks and spheroids on the basis of
the shape of the light profile parametrized by the Sersic index
$\mathrm{n}$. The contamination between the two classes is expected to
vary from $\sim8\%$ at $\mathrm{I=22}$ to $\sim17\%$ at
$\mathrm{I = 22.5}$.

\begin{description}
\item[{\it Colour -- magnitude distributions}.] Spheroids and disks
show an evident segregation in the CM diagrams. The first family of
galaxies define a sharp red sequence both in the optical and in the
NIR, while disks have in general bluer colours and are located
preferably below the CM relations. We find, however, some level of
mixing between the properties of the two classes: some spheroids have
blue colours, while some disks are red or redder than the red
sequences.  Interestingly, some of the blue spheroids are found at
bright magnitudes (I $< 20$) and are not expected to be
field contaminants.
\item[{\it Spatial distributions}.] 
Disks and spheroids show a sharp spatial segregation,
with disk galaxies found preferably in the low density regions.  The
spheroid population defines a central structure with a diameter of
$\sim1.0~Mpc$, and secondary clumps located at a distance $>1.0$ Mpc
from the main density peak. By analyzing the density map of blue and
red galaxies, we find that the two populations have a very different
spatial distribution, the blue galaxies showing an amorphous
structure.
\end{description}
A standard Butcher -- Oemler analysis has been performed in order to
derive the fraction of blue cluster galaxies. The cluster \cl0048 \,
is characterized by a concentration parameter $\mathrm{C\sim0.45}$,
that is intermediate between a very rich cluster and a less
concentrated, poor structure. The radius that encloses $30\%$ of the
total galaxy population is comprised between $1$ and $1.5$ arcmin
($0.4 - 0.6~\mathrm{Mpc}$). The fraction of blue galaxies amounts to
$\mathrm{0.11 \pm 0.07}$, and increases up to $\sim23\%$ at a
distance of $\sim1.6~\mathrm{Mpc}$.  We do not find, therefore, a
strong BO effect in the cluster \cl0048 at $\mathrm{z=0.64}$, in
agreement with previous studies who did not find a high increase of
the fraction of blue galaxies at $\mathrm{z > 0.5}$.

\begin{acknowledgements}
This work takes advantage of the VIRMOS photometric preparatory
survey.  We warmly thank O. Le F\`evre and the VIRMOS Consortium who
allowed us to use a subset of VIRMOS photometric data base to estimate
the field contribution.  In particular, we thank H. McCraken for the
help with the VIRMOS catalogues.  We are grateful to C. Lidman who
helped us for the calibration of the ISAAC photometry and to C. Lobo
for providing us with her spectroscopic sample. We thank the ESO staff
who effectively attended us during the observation run at FORS2. We
also thank M. Capaccioli and R. de Carvalho for the helpful
discussions and the anonymous referee for his/her comments which helped
us to improve the manuscript.  
Michele Massarotti is partly supported by a MIUR-COFIN grant.
\end{acknowledgements}


\begin{thebibliography}{}
\bibitem[1996]{ASH96} Abraham, R.G., Smecker-Hane, T.A., Hutchings, J.B.,
et al. 1996, \apj, 471, 694
\bibitem[1998]{AMK98} Adami, C., Mazure, A., Katgert, P., \& Biviano,
A. 1998, \aap, 336, 63
\bibitem[2002]{ACD02} Amico, P., Cuby, J.G., Devillard, N., Jung, Y.,
\& Lidman C. 2002, ISAAC Data Reduction Guide 1.5.
\bibitem[1999]{AnE99} Andreon, S., \& Ettori, S. 1999, \apj, 516, 647
\bibitem[1993]{AEC93} Arag\'on-Salamanca, A., Ellis, R.S., Couch,
W.J. \& Carter, D. 1993, \mnras, 279, 1
\bibitem[1996]{BAR96} Barger, A.J., Arag\'on-Salamaca, A., Ellis,
R.S., et al. 1996, \mnras, 279, 1
\bibitem[1990]{BFG90} Beers, T.C., Flynn, K., \& Gebhardt, K. 1990,
\aj, 100, 32
\bibitem[1996]{BeA96} Bertin, E., \& Arnouts, S. 1996, \aaps\, 117,
393
\bibitem[1993]{BrC93} Bruzual, G.A., \& Charlot, S. 1993, \apj, 405,
538
\bibitem[2002]{BML02} Busarello, G., Merluzzi, P., La Barbera, F., 
Massarotti, M., \& Capaccioli, M. 2002, \aap, 389, 787
\bibitem[1978a]{BO78a} Butcher, H., \& Oemler, A. 1978a, \apj, 219, 18
\bibitem[1978b]{BO78b} Butcher, H., \& Oemler, A. 1978b, \apj, 226, 559
\bibitem[1984]{BO84} Butcher, H., \& Oemler, A. 1984, \apj, 285, 426
\bibitem[1987]{CoS87} Couch, W.J., Sharples, R.M. 1987, \mnras, 229,
423
\bibitem[1998]{CBS98} Couch, W.J., Barger, A.J., Smail, I., Ellis,
R.S., \& Sharples, R.M. 1998, ApJ, 497, 188
\bibitem[1997]{DOC97} Dressler, A., Oemler, A., Jr., Couch, W.J.,
et al. 1997, \apj, 490, 577
\bibitem[2001]{ELY01} Ellingson, E., Lin, H., Yee, H.K.C., \& 
Carlberg, R.G. 2001, \apj, 547, 609
\bibitem[2002]{FJW02} Fairley, B.W., Jones, L.R., Wake, D.A., 
et al. 2002, \mnras, 330, 755
\bibitem[1999]{GAR99} Garilli, B., Maccagni, D., \& Andreon, S. 1999,
\aap, 342, 408
\bibitem[1998]{GLY98} Gladders, M.,D., Lopez-Cruz, O., Yee, H.K.C., \&
Kodama, T. 1998, \apj, 501, 571
\bibitem[1995]{HBC95} Holtzman, J.A., Burrows, C.J., Casertano, S.,
et al. 1995, \pasp, 107, 1065
\bibitem[1998]{HMT98} Hunt, L.K., Mannucci, F., Testi, L., et al.
1998, \aj, 115, 2594
\bibitem[1997]{KoA97} Kodama, T., \& Arimoto, N. 1997, \aap, 320, 41
\bibitem[1998]{KAB98} Kodama, T., Arimoto, N., Barger, A.J, \&
Arag\'on-Salamanca, A. 1998, \aap, 334, 99
\bibitem[2001]{KOB01} Kodama, T., \& Bower, R.G. 2001, \mnras, 2001,
312, 18
\bibitem[1988]{KKN88} Koo, D.C., Kron, R.G., Nanni, D.,
 Vignato, A., \& Trevese, D. 1988, \apj, 333, 586
\bibitem[1980]{KRO80} Kron, R.G. 1980, \apjs, 43, 305
\bibitem[2002]{LBM02} La Barbera, F., Busarello, G., Merluzzi, P.,
Massarotti, M., \& Capaccioli, M. 2002, \apj, 571, 790
\bibitem[1992]{LAN92} Landolt, A.U. 1992, \aj, 104, 340
\bibitem[2000]{LIL00} Lobo, C., Iovino, A., Lazzati, D., Chincarini,
G. 2000, \aap, 360, 896
\bibitem[2001]{MdC01} Margoniner, V.E., de Carvalho, R.R., Gal,
R.R., \& Djorgovski, S.G. 2001, \apj, 548, 143
\bibitem[2001]{MASSa} Massarotti, M., Iovino, A., \& Buzzoni, A. 2001a,
\aap, 368, 74
\bibitem[2001]{MASSb}Massarotti, M., Iovino, A., Buzzoni, A., \&
Valls-Gabaud 2001b, \aap, 380, 425
\bibitem[1995]{NFW95} Navarro, J.F., Frenk, C.S., \& White, S.D.M. 1995,
\mnras, 275, 720
\bibitem[1996]{NFW96} Navarro, J.F., Frenk, C.S., \& White, S.D.M. 1996,
\mnras, 462, 563
\bibitem[1999a]{OSdC99a} Olsen, L.F., Scodeggio, M., da Costa, L.,
et al. 1999a, \aap, 345, 363
\bibitem[1999b]{OSdC99b} Olsen, L.F., Scodeggio, M., da Costa, L.,
et al. 1999b, \aap, 345, 681
\bibitem[1998]{PMK98} Persson, S.E., Murphy, D.C., Krzeminski, W., Roth,
M., \& Rieke, M.J. 1998, \aj, 116, 2475
\bibitem[1998]{PIC98} Pickles, A.J. 1998, \pasp 110, 863
\bibitem[1997]{Pog97} Poggianti, B.M. 1997, \aaps, 122, 399
\bibitem[1996]{PoB96} Poggianti, B.M., \& Barbaro, G. 1996, \aap,
314, 379
\bibitem[1995]{RaS95} Rakos, K.D., \& Schombert, J.M. 1995, \apj, 439,
47
\bibitem[1997]{RdC97} Renzini, A., \& da Costa, L. 1997, Messanger 87,
23
\bibitem[2001]{SML01} Sabbey, C.N., McMahon, R.G., Lewis, J.R., \&
Irwin, M.J. 2001, Astronomical Data Analysis Software and Systems
X. In ASP Conf. Proceedings, ed. F. R. Harnden, Jr., F.  A. Primini,
\& H. E. Payne., vol. 238, 317
\bibitem[1986]{SCA86} Scalo, J.M. 1986, Fundamentals of Cosmic
Physics 11, 1
\bibitem[2001]{SLI01}Serote Roos, M., Lobo, C., \& Iovino, A. 2001,
Proceedings of the ESO/ECF/STScI Workshop Garkhing Germany 2000,
eds. S. Cristiani, A. Renzini, R.E. Williams. Springer 2001 p.215
\bibitem[1998]{SED98} Stanford, S.A., Eisenhardt, P.R.M., \&
Dickinson, M. 1998, \apj, 492, 461
\bibitem[1998]{vDF98} van Dokkum, P.G., Franx, M., Kelson, D.D.,
et al. 1998, \apj, 500, 714
\bibitem[2000]{vDF00} van Dokkum, P.G., Franx, M., Fabricant, D.,
Illingworth, G.D., \& Kelson, D.D. 2000, \apj, 541, 95
\end{thebibliography}
\end{document}